RESEARCH ARTICLE Open Access

# Mass campaigns with antimalarial drugs: a modelling comparison of artemether-lumefantrine and DHA-piperaquine with and without primaquine as tools for malaria control and elimination

Jaline Gerardin*, Philip Eckhoff and Edward A Wenger

## Abstract

**Background:** Antimalarial drugs are a powerful tool for malaria control and elimination. Artemisinin-based combination therapies (ACTs) can reduce transmission when widely distributed in a campaign setting. Modelling mass antimalarial campaigns can elucidate how to most effectively deploy drug-based interventions and quantitatively compare the effects of cure, prophylaxis, and transmission-blocking in suppressing parasite prevalence.

**Methods:** A previously established agent-based model that includes innate and adaptive immunity was used to simulate malaria infections and transmission. Pharmacokinetics of artemether, lumefantrine, dihydroartemisinin, piperaquine, and primaquine were modelled with a double-exponential distribution-elimination model including weight-dependent parameters and age-dependent dosing. Drug killing of asexual parasites and gametocytes was calibrated to clinical data. Mass distribution of ACTs and primaquine was simulated with seasonal mosquito dynamics at a range of transmission intensities.

**Results:** A single mass campaign with antimalarial drugs is insufficient to permanently reduce malaria prevalence when transmission is high. Current diagnostics are insufficiently sensitive to accurately identify asymptomatic infections, and mass-screen-and-treat campaigns are much less efficacious than mass drug administrations. Improving campaign coverage leads to decreased prevalence one month after the end of the campaign, while increasing compliance lengthens the duration of protection against reinfection. Use of a long-lasting prophylactic as part of a mass drug administration regimen confers the most benefit under conditions of high transmission and moderately high coverage. Addition of primaquine can reduce prevalence but exerts its largest effect when coupled with a long-lasting prophylactic.

**Conclusions:** Mass administration of antimalarial drugs can be a powerful tool to reduce prevalence for a few months post-campaign. A slow-decaying prophylactic administered with a parasite-clearing drug offers strong protection against reinfection, especially in highly endemic areas. Transmission-blocking drugs have only limited effects unless administered with a prophylactic under very high coverage.

## Background

Despite enormous reductions in malaria incidence and mortality in the past decade, malaria continues to pose a serious health risk to much of the world's population. In 2012, over 200 million cases and 600,000 deaths have been attributed to malaria [1]. As countries continue to implement control strategies and move from malaria control to elimination, it is crucial to understand how a variety of intervention methods are best deployed for maximum reduction in transmission.

Antimalarial drugs have been used for malaria control since the 1920s and were one of several tools employed in the eradication programs of the mid-twentieth century [2]. With the recent development of potent artemisinin-based combination therapies (ACTs), mass administration of antimalarial drugs is once again coming into play as one of many elements in plans for malaria control and elimination [3-5]. Although malaria is transmitted by vectors, the infectious reservoir is contained in humans. Thus, drug clearance of parasites within an infected population has the potential to interrupt transmission under the right circumstances, and under less optimal conditions drug-based campaigns may still reduce parasite prevalence for months.

* Correspondence: jgerardin@intven.com
Institute for Disease Modeling, Intellectual Ventures, 1555 132nd Ave NE, Bellevue, WA 98005, USA

In the field, mass drug administrations (MDAs) have met with mixed success. While prevalence is suppressed during and shortly after the MDA campaign, low prevalence often fails to be sustained 6 months after the end of campaigns, especially in regions of high endemicity [6]. In particular, transmission-blocking drugs such as primaquine appear to have minimal effects on prevalence [7,8].

Many questions remain regarding the determinants of campaign outcome. Is screening before treatment a viable alternative to mass administration given that current diagnostics have only limited sensitivity? Compliance with a complex drug regimen such as that of artemether-lumefantrine (AL) is estimated to be relatively low [9-11], but the effect of low compliance on population prevalence remains unknown. Prophylaxis is acknowledged as a powerful tool, but its importance may depend on local transmission intensity and campaign coverage, and in certain settings prophylaxis may only suppress prevalence by a negligible amount [2,12]. Primaquine is currently considered as a gametocytocide in treating *P. falciparum* [13,14], yet little is understood about the degree of prevalence reduction that can be gained by killing mature gametocytes. Because of the potential for very serious adverse events arising from treatment with primaquine [15,16], it is critical to understand whether mass administration with primaquine could significantly reduce prevalence.

Mathematical modelling has been used for many years as a tool for understanding the dynamics of malaria and for predicting the outcomes of interventions [12,17,18]. Recent work has built exceedingly sophisticated models capable of tracking the progress of individual infections, including the development of asexual parasites into gametocytes and the acquisition of host immunity [19-21]. Simulations allow the testing of many campaign scenarios in a wide range of settings, enabling detailed understanding of how campaign elements—coverage, timing, frequency, choice of drug—affect campaign outcome [22-27].

While several different models have been employed to predict outcomes of mass distribution of antimalarial drugs, none has modelled detailed simulation of drug pharmacokinetics (PK) and pharmacodynamics (PD) in a simulation of malaria transmission that includes within-host effects and mosquito dynamics. Here we implement PK models for artemether, lumefantrine, dihydroartemisinin, piperaquine, and primaquine in a previously established agent-based model of malaria transmission [19,20]. Asexual parasite and gametocyte killing effects are based on *in vitro* measurements of drug efficacy and calibrated to clinical outcomes [14,28-44]. Various campaign scenarios are compared: number of rounds of distribution in a campaign, mass-screen-and-treat versus mass administration, the influence of campaign coverage and compliance on campaign efficacy, usage of artemether-lumefantrine (AL) versus dihydroartemisinin-piperaquine (DP), and the effect of adding primaquine on transmission reduction.

## Methods

### Malaria transmission model

Simulations were conducted with EMOD v1.6 with a simulation timestep of 1 hour. The EMOD model of malaria transmission is a stochastic individual-based model with mosquito life cycle dynamics and species-specific feeding habits [19]. Infections begin with an infectious bite, and parasites progress through liver stage, asexual blood stage with antigenic variations, and 6 sexual stages. Each individual can sustain up to 3 simultaneous infections, and all parasite strains respond to antimalarial drugs with identical pharmacodynamics. Host immunity is modelled mechanistically and includes an innate response that clears asexual parasites and limits gametocyte success in mosquitoes, an adaptive response to variable epitopes that clears red blood cells infected with asexual parasites, and an adaptive response that limits the success of merozoite invasion [45]. Human infectiousness to mosquitoes depends on gametocyte density and human immune factors impacting gametocyte survivability within mosquitoes. No case management or vector control interventions were included. The EMOD software and documentation are available at http://idmod.org/software.

### Modelling pharmacokinetics (PK) of antimalarial drugs

We implemented a simplified PK for five antimalarials—artemether (AM), lumefantrine (LF), dihydroartemisinin (DHA), piperaquine (PPQ), and primaquine (PQ)—with a single or double exponential to model drug distribution and elimination (Figure 1). Parameters were calculated from compartmental model microconstants reported in the literature (Additional file 1: Table S1, Table S2) [22,28,29,46-59].

In the simple PK model, administration of drugs results in immediate absorption into the blood. Drug concentrations $C(t)$ decay with a double exponential to approximate 2-compartment pharmacokinetics:

$$C(t) = C_{max}\left(e^{-At} + Ve^{-Bt}\right) \quad (1)$$

Estimates of absorbance rate constants for these five antimalarials range from 0.083 to 2.18/h (Additional file 1: Table S2). Modeling absorption as immediate may result in overestimation of patients' exposure to certain drugs (LF and DHA), but the overexposure is small relative to the total exposure. Parameters in the double exponential model are derived from compartmental model parameters [60]. Maximum plasma drug concentration $C_{max}$ is given by

$$C_{max} = S\frac{dose}{V_c + V_p} \quad (2)$$

where $S$ is a scale factor chosen such that $C_{max}$ is close to maximum drug concentration observed in compartmental models, $V_c$ and $V_p$ are volumes of the central and peripheral compartments as described in compartmental models, and *dose* is the age-appropriate dose of drug. See Additional file 1: Table S1 for a complete list of compartmental parameter values used to construct the simple model.

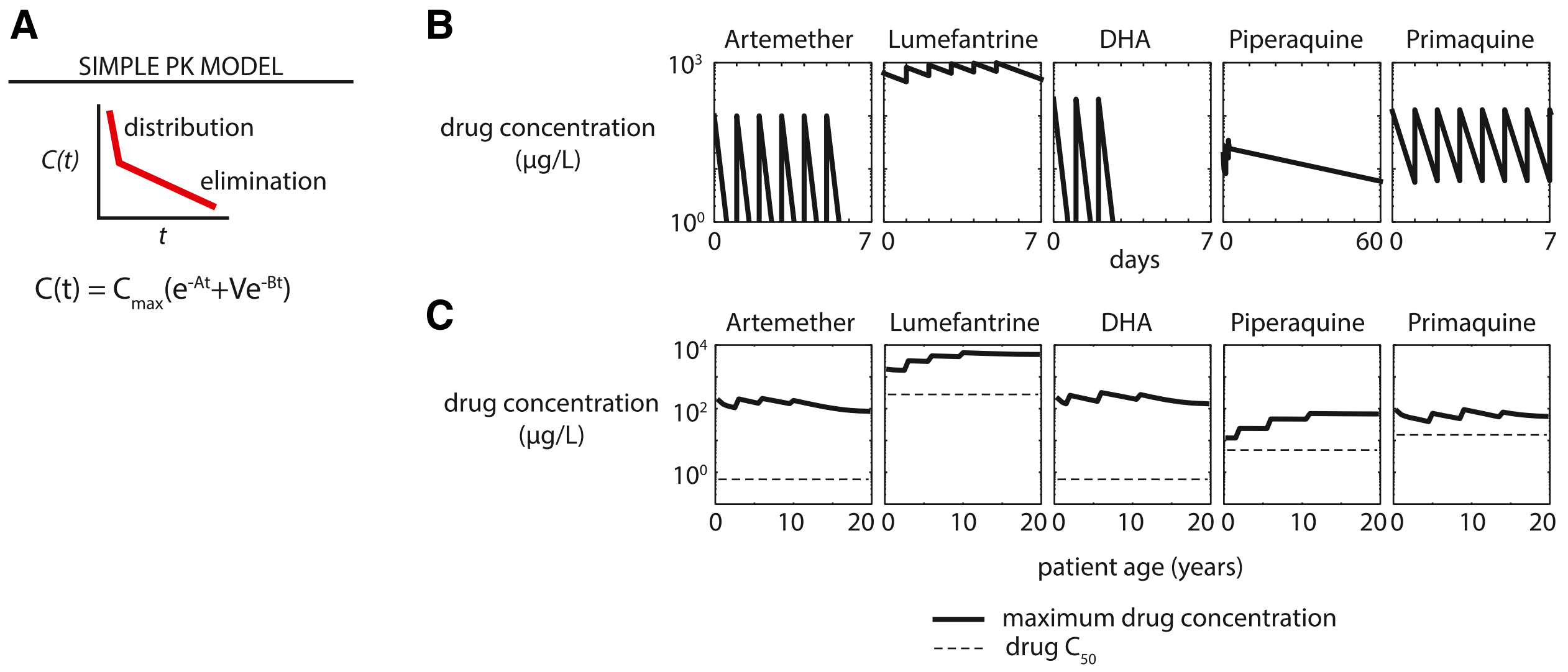

**Figure 1 Pharmacokinetics and pharmacodynamics of antimalarial drugs are implemented in an agent-based model of malaria transmission and calibrated to clinical data. (A)** A simplified model of drug pharmacokinetics uses a double exponential to model antimalarial drug distribution and elimination. **(B)** Pharmacokinetics of 5 antimalarials as simulated with the simple PK model in a 50 kg adult. **(C)** Maximum drug concentration with simple PK including age-based dosing, weight-dependent PK parameters, and WHO/CDC age-weight tables. See Tables 1 and 2 for model PK parameters and age-based dose fractions.

Decay time constants $A$ and $B$ are defined in terms of compartmental microconstants $k_{12}$, $k_{21}$, and $k_e$ as

$$A = \frac{1}{2}\left(k_{12} + k_{21} + k_e + \sqrt{(k_{12} + k_{21} + k_e)^2 - 4k_{21}k_e}\right) \quad (3)$$

$$B = \frac{1}{2}\left(k_{12} + k_{21} + k_e - \sqrt{(k_{12} + k_{21} + k_e)^2 - 4k_{21}k_e}\right) \quad (4)$$

and fractional volume $V$ as

$$V = \frac{k_e}{B} \quad (5)$$

For AM, DHA, and PQ, a single exponential decay was used to model drug PK. In the single exponential case, $V = 0$ and $A$ is the elimination rate from the compartmental model. Parameters used for modeling antimalarials in the simple exponential model are listed in Table 1.

AM, LF, DHA, and PPQ were administered as fixed-dose combinations, and PK of each drug was modeled independently. For each antimalarial drug, children received a fraction of the adult dose according to Table 2 [61]. AM and LF were given as the combination AL, consisting of 80 mg AM and 480 mg LF in the adult dose, taken every 12 hours over three days (6 doses total). DHA and PPQ were given as the combination DP, consisting of 120 mg DHA and 960 mg PPQ in the adult dose, taken once a day for three days (3 doses total). Single dose PQ was administered with a 15 mg adult dose, with children receiving a fractional dose according to Table 2.

**Table 1 Double exponential model PK parameters for 50 kg adult patient**

| | AM | LF | DHA | PPQ | PQ |
|---|---|---|---|---|---|
| **Adult dose** | **80 mg** | **480 mg** | **120 mg** | **960 mg** | **15 mg** |
| $C_{max}$ (μg/L) | 114 | 1017 | 200 | 30 | 75 |
| $C_{max}$ scaling factor | 0.5 | 1.5 | 0.25 | 0.5 | 1 |
| Decay constant A (1/day) | 0.12 | 1.3 | 0.12 | 0.17 | 0.36 |
| Decay constant B (1/day) | -- | 2.0 | -- | 41 | -- |
| V | -- | 1.2 | -- | 49 | -- |

**Table 2 Age-based dosing: children are given a fraction of the adult dose according to their age**

| AL (AM + LF) | | DP (DHA + PPQ) | | PQ | |
|---|---|---|---|---|---|
| Age | Dose fraction | Age | Dose fraction | Age | Dose fraction |
| 1 – 3y | 0.25 | 6 m – 2y | 0.17 | <5y | 0.17 |
| 3 – 6y | 0.5 | 2 – 6y | 0.33 | 5 – 9y | 0.33 |
| 6 – 10y | 0.75 | 6 – 11y | 0.67 | 9 – 14y | 0.67 |
| >10y | 1.0 | >11y | 1.0 | >14y | 1.0 |

### Calibrating pharmacodynamics of antimalarial drugs

Drug effects on parasite concentrations were modeled as a Hill function [27,62,63]:

$$kill\ rate = k_{max} \frac{C(t)}{C(t) + C_{50}} \tag{6}$$

$C_{50}$ of each drug was estimated from *in vitro* data from literature [29-33]. A Hill coefficient of 1 was assumed for all drugs. The maximum kill rate $k_{max}$ was stage-specific, while $C_{50}$ for each drug was the same for all parasite life cycle stages. Maximum kill rates for asexual parasites, gametocytes stages I to IIb, gametocytes stages III to IV, and mature gametocytes were taken from literature [29-33] when available and subsequently manually tuned to replicate parasite clearance times, recrudescence rates, reinfection rates, and gametocyte clearance times from clinical data [28,34-41]. See Table 3 for final calibrated $C_{50}$ values and stage-specific maximum kill rates. All final calibrated $C_{50}$ values fell within the range of *in vitro* measurements, although we find that $C_{50}$ for piperaquine must lie on the low end of *in vitro* observations in order to match clinical data for reinfection rates.

All pharmacodynamics calibrations were performed on a population of 1000 people of all ages with no births or deaths. Each simulation was repeated for 100 stochastic realizations.

To calibrate parasite clearance time, naïve and semi-immune patients were challenged in the absence of vectors with an infectious bite on day 0 and treated on day 25, shortly before the peak of asexual parasite density in the course of an untreated infection. Semi-immune patients were a population of individuals with the age-dependent immune systems of people living in an endemic region with annual entomological inoculation rate (EIR) 50. Parasite clearance time was defined as the number of days post treatment after which a patient's asexual parasitaemia fell below 10/μL. Maximum kill rate of asexual parasites was tuned for AM and DHA to achieve parasite clearance time of 1 day for most individuals [28,34-36]. Because AL is dosed twice a day while DP is dosed only once a day, resulting in more exposure to AM than DHA at current dosing levels, parasite clearance time tended to be slightly shorter for AL than for DP.

To calibrate recrudescence rates, semi-immune patients were challenged with an infectious bite on day 0 and treated on day 25 in the absence of vectors. Recrudescent patients were those with asexual parasitaemia above 10/μL on day 42 post-treatment (day 67 post-infection), and recrudescence rate was compared to reported values for ACTs [35-41].

To calibrate reinfection rates, semi-immune patients were subjected to a constant annual EIR of 36, approximating clinical trial conditions, for 1 year before treatment. Infected patients were those with asexual parasitaemia above 10/μL on day 42 post-treatment. The recrudescence rate was subtracted from the total infected rate to determine the rate of reinfection. The maximum kill rate of asexual parasites for LF and PPQ were tuned to achieve reinfection rates of 40% and 20% respectively at 42 days post-treatment [35-41].

Stage-specific gametocyte killing rates were initialized to *in vitro* measurements [64]. Semi-immune patients were challenged with an infectious bite on day 0, and treatment was administered to all patients with asexual parasitaemia greater than 10/μL on day 35, when gametocyte prevalence is highest in an untreated population. Gametocyte prevalence was measured on days 0, 7, 10, and 14 post treatment. Gametocyte kill rates of AM were tuned to achieve gametocyte clearance in 7–10 days [14,42,43], and gametocyte kill rates for DHA were assumed to be the same as those for AM. A small immature gametocyte killing effect was added to LF and PPQ in line with *in vitro* observations.

To calibrate PQ inactivation of gametocytes, each treated patient's infectiousness toward mosquitoes was measured 2 days post treatment. PQ kill rate of mature

**Table 3 Model parameters for pharmacodynamics**

| | AM | LF | DHA | PPQ | PQ |
|---|---|---|---|---|---|
| $C_{50}$ (μg/L) | 0.6 | 280 | 0.6 | 5 | 15 |
| Asexual parasite maximum kill rate (1/day) | 8.9 | 4.8 | 9.2 | 4.6 | 0 |
| Gametocyte stage I-IIb maximum kill rate (1/day) | 2.5 | 2.4 | 2.5 | 2.3 | 2.0 |
| Gametocyte stage III-IV maximum kill rate (1/day) | 1.5 | 0 | 1.5 | 0 | 5.0 |
| Mature gametocyte maximum kill rate (1/day) | 0.7 | 0 | 0.7 | 0 | 50.0 |
| Hepatocyte stage maximum kill rate (1/day) | 0 | 0 | 0 | 0 | 0 |

gametocytes was tuned to achieve high levels of inactivation for the 0.1 mg/kg dose [44].

### Simulating mass campaigns with antimalarial drugs

Seasonal temperature, rainfall, vectors, and larval habitat abundance were modelled on the Zambia Sinazongwe site. Transmission between humans and vectors resulted in an average annual EIR of 50 in the absence of interventions with age-specific immunity corresponding to that level of transmission. Births and deaths were allowed such that the population remained around 1000 individuals. Populations were calibrated with a 2-year burn-in period prior to any interventions. Campaigns were conducted during year 3. All interventions were administered on the same day for all people. Drug distribution rounds of the three-round campaign occurred on days 170, 226, and 282 of year 3, approximating six weeks of distribution in the field interspersed with two weeks out of the field. The two-round campaign rounds occurred on days 250 and 292, approximating a four-week distribution schedule with two weeks in between. Day 324 marked the end of campaigns, and all dates referencing the end of campaign are relative to this date. Prevalence was calculated based on an asexual parasite detection threshold of 10/μL unless otherwise indicated. All simulations were repeated for 100 stochastic realizations.

Annual EIRs lower than 50 were simulated by reducing available larval habitat as indicated in Additional file 1: Table S3. Population immunity was initialized by 50-year burn-in with the appropriate EIR prior to the 2-year burn-in described above.

Coverage for each individual was independent in each campaign round, and no group of individuals was systematically missed for all rounds. All individuals reached by campaign took at least one directly observed dose of drugs. Subsequent doses were taken based on a random draw against the compliance parameter. For example, at 20% compliance, each dose after the first had 20% chance of being taken. In a multidrug regimen, each drug was complied with separately.

### Quantification of effects of primaquine

Three-round MDA campaigns were simulated with administration of AL, AL + PQ, DP, and DP + PQ according to the schedule described above for coverage between 50% and 100% and EIRs between 1 and 50 for each drug and coverage level. Each EIR, coverage, and drug combination was repeated for 100 stochastic realizations and prevalence was measured 4 months after the end of campaigns with an asexual parasite detection threshold of 0.05 parasites/μL. A highly sensitive detection threshold was chosen to distinguish between very low prevalence levels, which can be observed in scenarios with high coverage or low EIR.

Stochastic realizations resulting in local elimination, defined as zero prevalence for days 630 through 730 (100 days beginning 306 days after the end of campaign), were removed prior to subsequent analysis. Mean prevalence was bootstrapped with 1000 resamples of size 100. For each ACT and coverage level, prevalence means at all EIR levels for ACT alone and ACT + PQ were correlated using the Python 2.7 polyfit function with degree 1. The relationship between prevalence with ACT alone and ACT + PQ is expected to be linear at low EIR but not at high EIR. We are considering only lower EIRs and do not observe bowing in the plot of prevalence with ACT alone vs prevalence with ACT + PQ (Additional file 2: Figure S3). The relative prevalence reduction upon addition of PQ was calculated as 1 minus the fitted slope. Slopes were fitted to each resampled mean prevalence pair (ACT and ACT + PQ) at constant coverage level, allowing calculation of mean relative prevalence reduction upon addition of PQ at every coverage level for each ACT.

## Results

### Under-dosing of ACTs in children

Since pharmacokinetics are affected by body weight, we expect maximum plasma drug concentrations to vary according to both patient age and weight. Antimalarial drugs are administered to children under dosing regimens determined by age (Table 2). Age-dependence of drug concentrations will vary between drugs according to the nature of each drug's pharmacokinetic dependence on body weight.

Using age-weight charts from the CDC and WHO, we show in Figure 1C that age-based dosing achieves approximately uniform maximum plasma drug concentrations for AM, DHA, and PQ. While lower maximum concentrations of LF are observed in younger children, patients of all ages are able to achieve maximum LF concentrations above LF's estimated C50 of 280 μg/L. In contrast, under current dosing recommendations children below the age of 2 have maximum PPQ concentration much closer PPQ's estimated C50 of 5 μg/L; prophylactic effects of PPQ will therefore have much shorter duration. The under-dosing of children with piperaquine under current dosing recommendations has been well-documented in literature [65], but as amended guidelines have yet to be released, we have chosen to use the under-dosed treatment recommendations in our simulations.

Pharmacodynamics of the combination therapies AL and DP as well as gametocyte killing of PQ were tuned to match parasite clearance times, recrudescence rates, and reinfection rates to clinical data (Figure 2). As predicted by the age- and weight-based models of maximum drug concentrations, children treated with DP had higher recrudescence and reinfection rates than adults did due to PPQ levels insufficient for cure and prophylaxis (Figure 2C). Young children were especially vulnerable to recrudescence, while all infections were cured in adults, indicating that piperaquine was failing to kill the remaining parasites untouched by DHA. Likelihood of reinfection decreased with age among individuals treated with DP. Field studies have also shown that DP is likely under-dosed in children, and a higher dosage has been suggested [65-67]. In contrast, young children receiving AL were not more likely to be reinfected than adults [68].

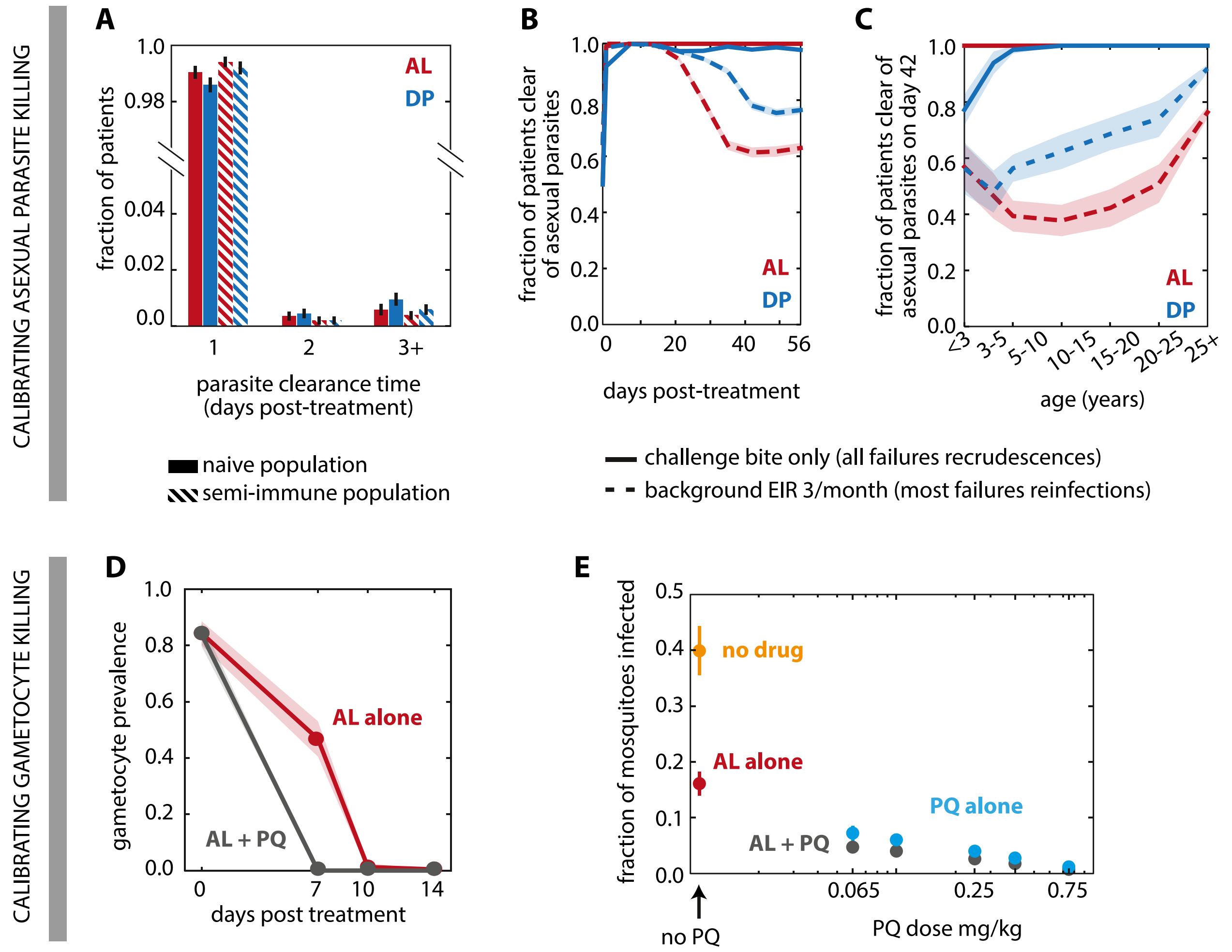

**Figure 2 Parasite killing rates of antimalarial drugs are calibrated to clearance time, prevalence, and infectivity. (A)** Asexual parasite clearance time is within 3 days post-treatment for AL (red) and DP (blue) in both naive (solid bar) and semi-immune (hashed bar) populations. Black bars: 95% confidence interval. **(B)** AL (red) and DP (blue) pharmacodynamics are calibrated in a semi-immune population to <3% recrudescence rate (solid lines) at day 42 post-treatment and 50% and 20% reinfection rate (dashed lines) at day 42 post-treatment with a background EIR of 3/month. Shaded areas: 95% confidence interval. **(C)** Fraction of patients clear of parasites at day 42 post-treatment is dependent on patient age for DP-treated population (blue) but not AL-treated population (red). Young children are more likely to experience recrudescence (solid line) and reinfection (dashed line) than older patients when treated with DP. Shaded areas: 95% confidence interval. **(D)** Gametocyte clearance time for AL-treated patients of all ages is calibrated to 7–10 days (red lines) post-treatment. Addition of single-dose PQ at 0.75 mg/kg reduces gametocyte clearance time to < 7 days. Shaded areas: 95% confidence interval. **(E)** Mean fraction of mosquitoes infected on day 2 post-treatment with AL, PQ, AL + PQ, or no drug in a semi-immune population challenged with an infectious bite 25 days prior to treatment. Primaquine was given with a single dose at 0.065, 0.10, 0.25, 0.40, and 0.75 mg/kg. Bars: 95% confidence interval (smaller than dot radius for PQ alone and AL + PQ).

### Interpretation of PQ pharmacodynamics

Experimental work suggests that PQ may render mature gametocytes incapable of infecting mosquitoes even while the gametocytes remain in the host bloodstream [13]. In our model, we are most interested in PQ's transmission-blocking activity and thus do not explicitly model inactivated gametocytes, instead considering them to be killed and cleared. We find that to replicate the observation that even a low PQ dose of 0.065 mg/kg PQ results in very few infected mosquitoes 2 days post treatment when given with a full course of AL (Figure 2E) [44], the maximum kill rate of PQ on mature gametocytes must be very high in order to compensate for PQ's short half-life of 8 hours. As currently parameterized, PQ completely clears gametocytes (Additional file 2: Figure S1D), making our predictions of its impact on prevalence an upper bound on its actual effect. However, incomplete knowledge of PQ's active metabolites and their pharmacokinetic profiles means that our model of PQ's killing action is only a gross approximation, and further refinement will be necessary as we learn more about PQ.

### ACTs in mass drug campaigns: comparison of AL and DP, 1 month post campaign

We test the effects of distributing AL and DP in a mass administration context. Case management and vector control were excluded in order to focus explicitly on the effects of the drug campaigns. Using an isolated population experiencing seasonal transmission, a multi-round drug administration campaign was simulated during the dry season, when prevalence drops to 30%. Both AL and DP reduce prevalence while the campaign rounds are ongoing and continue suppressing prevalence for a few months into the high transmission season (Figure 3A). Due to PPQ's long prophylactic tail, campaigns with DP result in lower prevalence that extends longer into the high transmission season than campaigns with AL as has been previously observed with generic short- and long-acting ACTs [25]. At this level of EIR, prevalence returns to baseline levels within 18 months of the MDA campaign even when DP is used. Using antimalarials to reduce prevalence in the long term requires repeated mass drug campaigns, permanent scale-up of case management with effective drugs, or simultaneous deployment of vector control with high coverage [12,23-25]. Here we focus on one aspect of a complex malaria control strategy: how to optimize elements of a single drug campaign to reduce parasite prevalence up to 4 months after the end of the campaign.

### ACTs in mass drug campaigns: comparison of 3-round and 2-round campaigns, 1 month post campaign

When malaria prevalence is seasonal, mass administration campaigns are conducted during the low transmission season in order to most effectively deplete the infectious reservoir and possibly interrupt transmission [23,69]. We compared the efficacy of three-round and two-round mass drug administration (MDA) campaigns at reducing prevalence one month after the end of the campaign. While 3-round campaigns cost more than 2-round campaigns, the third round offers an additional chance to reach people who were not covered in previous rounds. At a coverage level of 70%, and independent coverage between rounds, only 9% of individuals will have never been treated. We investigate how critical it is to reach another 70% of the last 9%.

We find that reducing campaign rounds from 3 to 2 increases parasite prevalence one month post-campaign for both AL and DP (Figure 3B). DP is more sensitive to number of rounds than AL. At 70% coverage, prevalence one month post-campaign is more than twice as high for a 2-round campaign as a 3-round campaign when DP is administered. Because DP has long-lasting prophylactic effects, additional campaign rounds are particularly beneficial as previously uncovered individuals become protected against reinfection and prevalence reduction is cumulative from round to round. In contrast, individuals treated with AL are not protected from reinfection for as long, and to some extent prevalence is able to reset to higher levels between rounds, leading to a smaller cost of reducing a campaign to fewer rounds.

When coverage is 100%, prevalence continues to be higher for the 2-round campaign, but the difference between 2- and 3-round prevalence is not as great (Additional file 2: Figure S2A). Switching from a 3-round campaign structure to a 2-round campaign is therefore recommended only for situations where very high coverage can be achieved. In all cases, treatment with DP showed lower prevalence post-campaign than treatment with AL.

### ACTs in mass drug campaigns: comparison of MSAT and MDA, 1 month post campaign

Campaigns may distribute antimalarial drugs to all individuals (mass drug administration, MDA), or they may choose to give drugs only to individuals testing positive for parasites (mass screen-and-treat, MSAT). If an insensitive diagnostic is used in an MSAT to identify individuals carrying parasites, then the MSAT campaign will fail to eliminate a large portion of the parasite reservoir, as many individuals who harbour sub-patent infections will fail to receive treatment. However, MSATs may be preferred because they avoid unnecessary dosing of uninfected individuals and may be less likely to lead to drug resistance in parasites.

Current rapid diagnostic tests (RDTs) are sensitive only above 50–200 parasites/μL [70], which has been shown to be inadequate for reducing prevalence in a mass campaign context [71-73]. To identify a minimum diagnostic sensitivity necessary for an MSAT campaign to reduce parasite prevalence with efficacy comparable to an MDA, we tested MSAT screening sensitivities from 0.01 to 200 parasites/μL (Figure 3C). Only diagnostics capable of detecting parasites below 0.1 parasites/μL result in prevalence reduction on par with an MDA campaign; current RDTs are nowhere near sensitive enough and new technologies are necessary if MSATs are to become the campaign of choice in the future. For DP, even MSATs with sensitivity of 0.01 parasites/μL cannot achieve the degree of prevalence suppression seen in an MDA because treating uninfected individuals with DP confers protection against infection. In contrast, AL has little prophylactic effect, and an AL MSAT with a highly sensitive diagnostic can be a good alternative to an AL MDA.

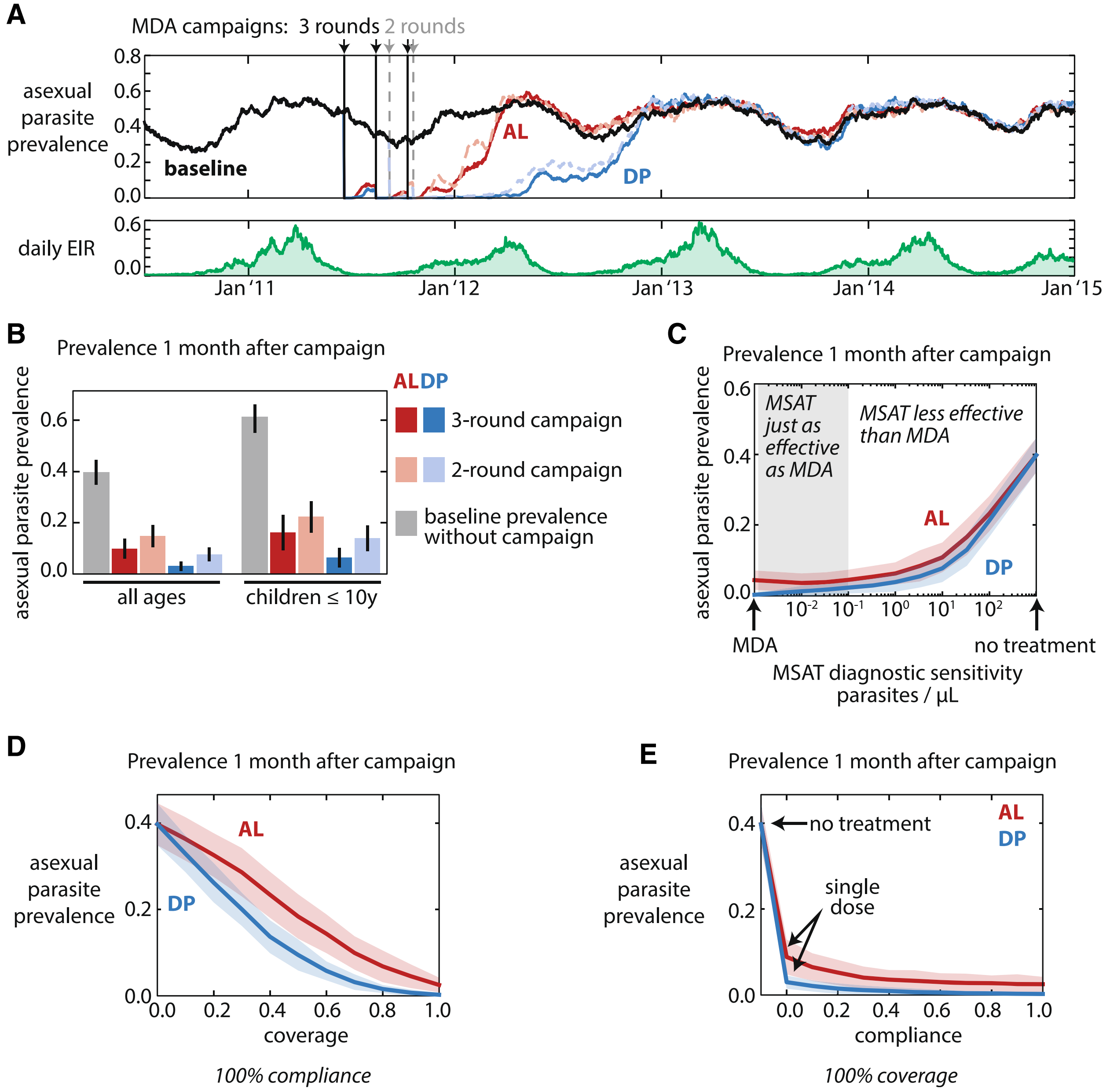

**Figure 3 Campaign outcome depends on timing, diagnostic sensitivity, coverage, and compliance.** **(A)** Asexual parasite prevalence (top) and daily EIR (bottom) of a semi-immune population of 1000 people with no intervention (black), multi-round MDA with AL (red), and multi-round MDA with DP (blue). Solid lines: 3-round campaigns. Dashed lines: 2-round campaigns. MDAs were simulated with 100% coverage and 100% compliance. Annual EIR was 50, and infected individuals were those with ≥10 asexual parasites/μL. DP's long prophylactic tail better protects against reinfection after campaign. **(B)** Prevalence 1 month after campaign for 3-round and 2-round MDA campaigns with AL or DP. MDAs were simulated with 70% coverage and 100% compliance; annual EIR was 50. Error bars: 95% confidence interval. **(C)** Prevalence 1 month after 3-round MSAT campaigns with varying sensitivity of MSAT diagnostic. Coverage and compliance were 100%; annual EIR was 50. Shaded areas: 95% confidence interval. **(D)** Prevalence 1 month after 3-round MDA campaigns with varying coverage and 100% compliance. Annual EIR was 50. Shaded areas: 95% confidence interval. **(E)** Prevalence 1 month after 3-round MDA campaigns with varying compliance and 100% coverage. All covered individuals take the first dose of AL or DP; subsequent doses are taken with probability equal to the compliance. Annual EIR was 50. Shaded areas: 95% confidence interval.

### ACTs in mass drug campaigns: effects of coverage and compliance, 1 month post campaign

Success of an MDA campaign may depend on both coverage, the fraction of the population reached by the campaign,

and compliance, individuals' rate of adherence to the correct drug dosing regimen. We tested the influence of coverage and compliance rates on outcomes of campaigns with AL and DP by comparing asexual parasite prevalence one month after the end of campaign rounds.

Increasing coverage up to 90% results in reduced prevalence for both AL and DP campaigns (Figure 3D). Under high transmission conditions with annual EIR of 50, each additional 20% increase in coverage reduces prevalence by half. Beyond 90%, further increase in coverage nets little additional gain. Campaigns deploying DP result in lower prevalence than campaigns using AL. To achieve the same prevalence reduction one month post-campaign, an AL campaign requires coverage of 15-20% higher than the DP campaign.

A sweep over compliance rates shows that most of the reduction in parasite prevalence is accomplished by taking the first dose of AL or DP (Figure 3E). In a semi-immune population, most infected individuals have low parasitaemia (Additional file 2: Figure S1), and a sub-curative dose of ACT is often capable of curing infections with low parasite density. Increasing compliance with unobserved doses from 10% to 100% reduces prevalence only by around 5%.

Coverage and compliance do not complement each other, and increasing compliance can compensate for deficiencies in coverage only to a very limited extent (Additional file 2: Figure S2C). For both AL and DP, increasing coverage results in stronger prevalence reduction than a similar increase in compliance. Compliance exerts a stronger effect on prevalence 4 months post-campaign for DP (Additional file 2: Figure S2D), when improving very low compliance for high coverage campaigns can reduce prevalence by 20% when compliance is increased from 0% to 20%. Improving compliance with DP campaigns lengthens the duration of protection against reinfection, which is a critical component of DP's efficacy in suppressing prevalence.

### Prophylaxis is most beneficial when transmission is high and coverage is moderately high: comparison of AL and DP, 4 months post campaign

In the simulated campaigns discussed above, administration of DP results in equal or lower prevalence than administration of AL under identical conditions. DP and AL differ primarily in the choice of partner drug to the artemisinin-based component. While lumefantrine is the more effective killer of asexual parasites, piperaquine possesses a much longer half-life and therefore confers a longer window of protection against reinfection.

The power of DP's long prophylactic tail is most striking when considering asexual parasite prevalence four months post-campaign in a high transmission setting (Additional file 2: Figure S2B). A DP campaign with only 45% coverage achieves on average the same prevalence reduction as an AL campaign with 100% coverage at this timepoint. Difficulties achieving good coverage can be overcome by choosing an antimalarial with superior prophylactic qualities.

To quantify the effect of a long-lasting prophylactic such as piperaquine, we assumed that all prevalence reduction observed after a campaign with AL is due to curing individuals, while any additional reduction in prevalence achieved by DP compared to AL is due to protection against reinfection (Figure 4A). Since both AL and DP cure nearly all individuals when correctly administered in the absence of reinfection (Figure 2), this assumption is reasonably accurate.

In a highly endemic setting, prophylaxis can suppress prevalence far beyond what can be achieved by cure alone (Figure 4B left). Even after an MDA campaign with good coverage, the infectious reservoir remains substantial and treated individuals remain vulnerable to reinfection, so prophylaxis is very powerful. Addition of a transmission-blocking drug such as PQ to an AL- or DP-based campaign further reduces prevalence, but prophylaxis has the larger effect.

Under low transmission conditions, administration with AL is able to deplete the infectious reservoir through cure. Because very few infected individuals remain after an MDA campaign with good coverage in a region with low transmission, protection against reinfection post-campaign offers little additional benefit (Figure 4B right). In contrast to high transmission settings, transmission-blocking drugs are approximately as effective at reducing transmission as prophylactics when coupled with a curative antimalarial in a low transmission setting.

The dependence of prevalence reduction due to prophylaxis on both coverage and EIR is shown in Figure 4C by measuring the difference in prevalence 4 months after MDA campaigns using AL and DP. As discussed above, prophylaxis reduces prevalence most for high EIR, where risk of reinfection is highest and protection against reinfection confers the most benefit. For coverage between 50 and 80%, prophylaxis reduces prevalence by an increasing amount as more individuals are protected against reinfection. For coverage above 80%, prophylaxis offers less additional gain in prevalence reduction because the infectious reservoir is already depleted through cure. Thus, each individual is challenged with few infectious bites, and prophylactic benefits are smaller.

### Addition of primaquine to MDA with ACTs confers a small additional reduction in prevalence: comparison of AL and DP with and without PQ, 4 months post campaign under variable EIR

Current WHO guidelines recommend the addition of primaquine (PQ) to an ACT regimen as a gametocytocide against *P. falciparum* infections [61]. To systematically quantify the impact of PQ on ACT campaigns, we compared prevalence four months after MDA campaigns with and without PQ over a range of transmission intensities and coverage levels. While lowering EIR is not a perfect proxy for altered conditions under vector control, it is an acceptable approximation for our study of relatively short-term outcomes. Our results agree with previous modelling observations that single-dose PQ as part of an MDA reduces prevalence only a small amount beyond that achieved by ACTs alone [23,25,27].

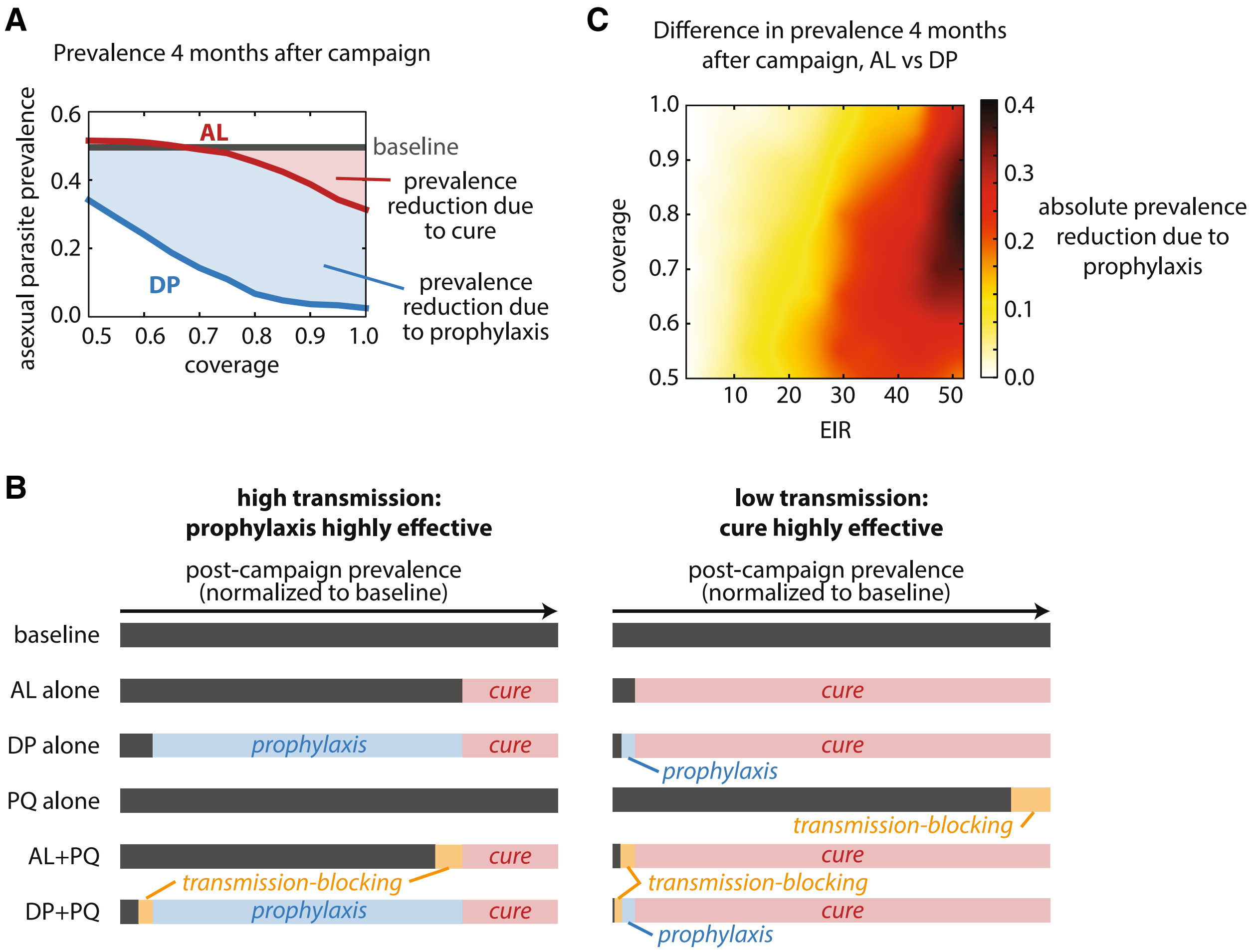

**Figure 4 Efficacy of prophylaxis is greatest under high transmission conditions and moderately high coverage. (A)** AL reduces prevalence by curing individuals in an infected population, while additional prevalence reduction observed in DP is due to the long-lasting prophylactic effects of piperaquine. Annual EIR of 50, MDA coverage 90%, compliance 100%, 3 campaign rounds, mean of 100 stochastic realizations. **(B)** Normalized prevalence 4 months after MDA campaigns in high and low transmission settings. Transmission-blocking refers to prevalence reduction upon addition of PQ. Coverage 90%, compliance 100%, 3 campaign rounds, annual EIR of 50 (left) and 1 (right). **(C)** Absolute prevalence reduction due to prophylaxis (mean prevalence using AL - mean prevalence using DP) depends on both coverage and EIR.

We can measure the relative effect of PQ by calculating the slope of a linear regression between prevalence outcomes for campaigns with and without PQ at constant coverage over a range of EIRs (Figure 5A) (see Methods). For the example in Figure 5A, this relative reduction is 0.025, indicating that at any level of annual EIR in our sampled range, addition of PQ to a campaign with AL that has 70% coverage will, on average, reduce prevalence at the 4-month mark by 2.5% relative to the prevalence that would have been observed had only AL been deployed. See Additional file 2: Figure S3 and Additional file 1: Table S4 for prevalence correlations at all coverage levels for both AL and DP.

Higher coverage results in PQ reducing prevalence by a higher relative amount for both AL and DP campaigns (Figure 5B). For AL, complete coverage of 100% yields a relative prevalence reduction of 13% when PQ is present. Thus, we expect that for mass drug campaigns that do not employ a long-lasting prophylactic, dosing with PQ can further reduce prevalence by only 13% or so. Because we did not account for some individuals' inability to be administered PQ or metabolize PQ to its active product, we expect 13% to approximate an upper bound on PQ impact. However, even at 100% coverage, MDA with AL results in prevalence of 30% at the 4-month mark, so a 13% additional relative reduction in prevalence can still reduce absolute prevalence by 4% overall.

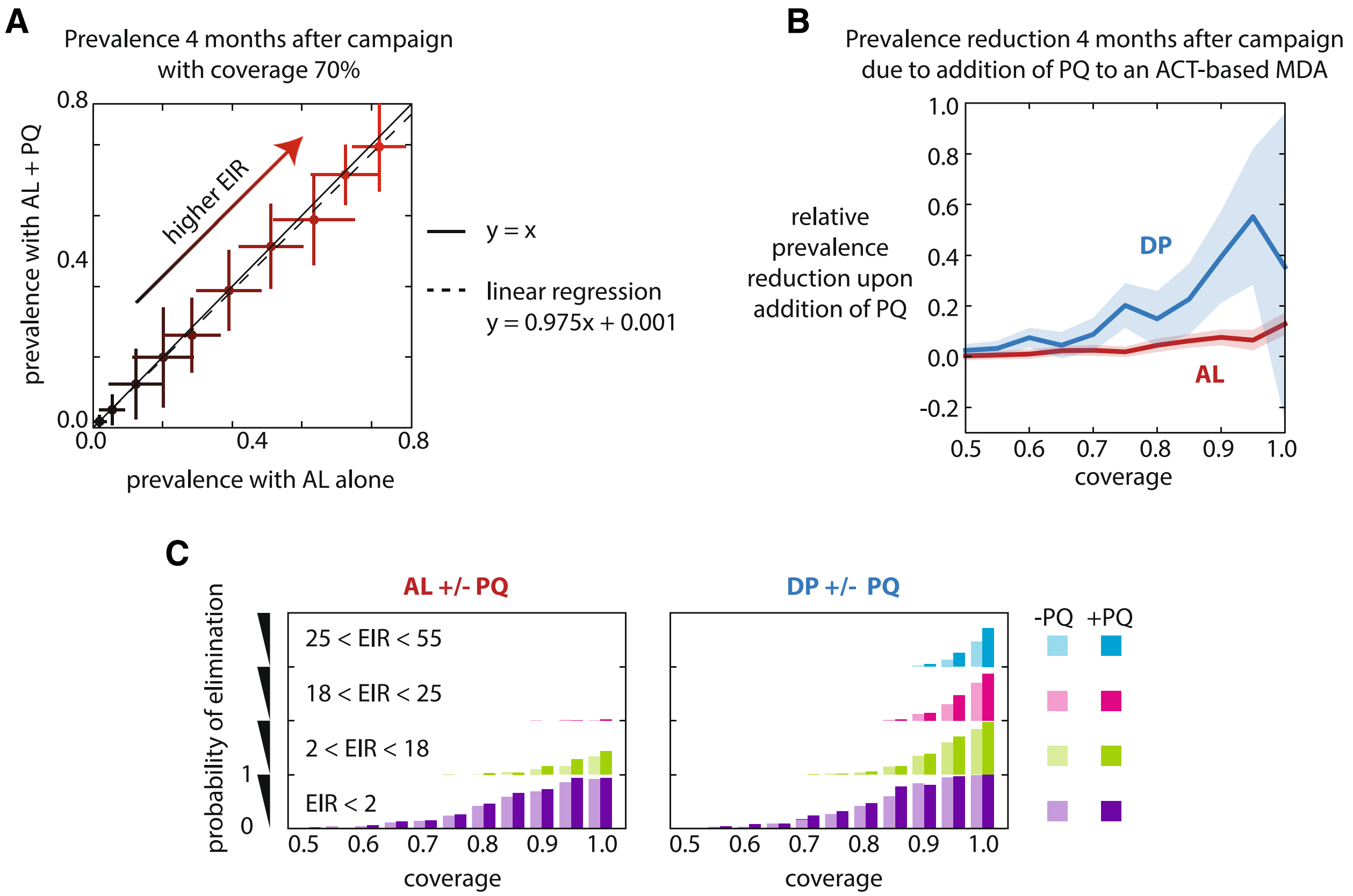

**Figure 5 Primaquine reduces transmission when campaign coverage is very high. (A)** At a coverage of 70%, addition of PQ to an AL MDA reduces prevalence 4 months post campaign by <5%. EIR is sampled at constant coverage for 100 stochastic realizations for each EIR. Compliance was 100% and 3 rounds of campaigns were conducted. Mean +/− one standard deviation is shown for each EIR value. A linear regression is shown (dotted line) correlating prevalence after MDA with AL + PQ to prevalence after a 3-round MDA with AL alone. **(B)** Prevalence reduction due to addition of PQ is highly dependent on coverage. The fractional relative prevalence reduction upon addition of PQ is the slope of the linear regression of prevalence with PQ to prevalence without PQ (panel A) subtracted from 1. Lines: bootstrapped mean fractional relative prevalence reduction at each level of coverage. Shaded areas: 95% confidence interval. **(C)** Higher coverage with ACTs and addition of primaquine lead to higher probability of local elimination, especially at higher EIR. Fraction of stochastic realizations leading to eradication is shown for 4 bins of annual EIR at each level of coverage, with 100 realizations per EIR per coverage. Compliance was 100% and 3 rounds of campaigns were conducted. The height of each bar is normalized to the total number of realizations in each EIR bin.

When coverage is sufficiently high, a MDA with DP alone can result in very low prevalence 4 months after the end of campaign. Most individuals are protected against reinfection for many weeks after their last round of DP, and only those who were not reached by MDA and those experiencing drug failure are vulnerable. Addition of PQ to the MDA can be very powerful in this context: PQ reduces prevalence another 60% relative to what was achieved by DP alone. By clearing mature gametocytes, PQ reduces the infectious period of an ACT-treated gametocyte carrier from 7 days to only 1 day. Thus, most individuals in the DP + PQ campaign are both non-infectious and protected against reinfection, leaving only the small fraction of uncovered individuals to propagate the parasite.

Increasing coverage increases the likelihood of local elimination for AL- and DP-based campaigns at all sampled EIR levels (Figure 5C). At low EIR and high coverage, ACTs alone eliminate malaria in nearly all simulations, but only DP can eliminate when transmission is high. Addition of PQ increases the likelihood of elimination but exerts its strongest effect under high EIR when deployed in combination with DP with high coverage.

## Discussion

Mass drug administrations are currently being considered as a tool for malaria elimination where vector control and case management have already reduced the disease burden. Here we investigate the conditions under which mass drug campaigns can have their strongest effects.

Drug campaigns lower malaria prevalence in three ways: by reducing the parasite reservoir in a population, by preventing infection of uninfected people, and by preventing transmission from infected people. High cure rate, high coverage, good compliance, multiple campaign rounds, and targeting a high number of low-density infections all contribute to reducing the parasite reservoir. While we have modelled coverage and compliance as independent between rounds, in reality it is likely that some individuals will be repeatedly missed by the campaign, refuse treatment, or display poor compliance with dosing schedules. Because compliance appears to have little effect on post-campaign prevalence, correlated compliance is unlikely to have major impact on modelled outcomes. As a rough example of the importance of avoiding significant missed pockets of transmission, compare a 3-round campaign with coverage of 70%. If coverage is perfectly correlated, the 3 campaign rounds result in overall coverage of 70%. With independent sample, coverage of 70% over 3 rounds is roughly equivalent to a case where only 3% of the population never receives drug, or a correlated coverage of 97%. Thus the modelled outcomes for a given coverage level represent a best-case scenario, and any correlation in untreated individuals across rounds will result in higher parasite prevalence and decrease the likelihood of elimination.

Population movement is also a crucial factor in near-elimination scenarios. Human movement around the campaign area such that some people are not at home during campaigns can be approximated with lower coverage. The reintroduction of malaria from people carrying infections into the campaign area and the robustness of elimination to reintroductions are critical to consider when modelling near elimination and will be addressed in subsequent work.

Prophylaxis, the prevention of infection in uninfected people, is most beneficial under high transmission conditions when individuals are frequently challenged with infectious bites. However, mass drug campaigns are unlikely to be conducted under such conditions. Under low or transmission conditions, biting rate is low, and high coverage with a prophylactic drug makes the infectious bite of a susceptible individual an unlikely event, pushing the system toward elimination.

Transmission-blocking and prophylaxis work together to reduce prevalence and increase the likelihood of elimination. On the whole, adding a transmission-blocking drug such as PQ to a mass drug campaign only reduces prevalence by a small amount. However, PQ should be considered in certain conditions: if the MDA campaign includes a long-lasting prophylactic, if coverage is very high, and if EIR is low and local elimination is the goal. Outside of these conditions, the risks and costs of mass distribution of PQ may very well outweigh the benefits.

## Conclusions

We developed an age- and weight-based model of antimalarial drug pharmacokinetics in the context of an agent-based model of malaria transmission. Using current drug dosing guidelines and available clinical data, our model predicts that young children are likely to be under-dosed for DHA-piperaquine, leaving them disproportionately vulnerable to recrudescence and reinfection.

Mass distribution of ACTs can reduce parasite prevalence for several months after the campaign. Mass-screen-and-treat campaigns administer antimalarials only to individuals who test positive for asexual parasites, but poor sensitivity in diagnostic tools means that individuals with very low parasitaemia, who may be infectious, remain untreated. Our model predicts that a diagnostic must achieve sensitivity at or below 0.1 parasites/μL in order for a mass-screen-and-treat campaign to have outcome similar to a mass drug administration.

By sweeping over possible coverage and compliance levels, we show that coverage exerts a much stronger effect on campaign outcome than compliance does. Late in the dry season, many individuals in a highly endemic area have asexual parasite densities low enough to be cleared by a single dose of ACT, so increasing compliance to complete an entire drug regimen has little impact on transmission. In contrast, increasing coverage to clear more infections will have a larger impact on the population's infectious reservoir.

A long-lasting prophylactic such as piperaquine confers protection against reinfection and is most efficacious for regions with high transmission and campaigns with moderately high coverage. For low transmission regions, changing policy to favour DP over AL offers much smaller benefits because risk of reinfection is lower.

Current interest in primaquine as a gametocytocide is very high. We find that single-dose primaquine is most efficacious when deployed with a long-lasting prophylactic like DP, where most of the population is protected against reinfection and only a small number of individuals are either infected or vulnerable to infection. For conditions where coverage is insufficiently high, prophylaxis is absent, or transmission rate is high, we predict that there is negligible benefit to adding primaquine to an MDA campaign.

Our simulations describe the efficacy of antimalarial drugs when deployed in a campaign setting. We anticipate that the addition of vector control and vaccines may interplay with MDAs in interesting ways as prevalence is reduced close to 0, and future work will situate drug campaigns in the context of other ongoing interventions. We anticipate that after other interventions have reduced transmission intensity, MDAs with high coverage can be a very powerful tool for malaria control and elimination.

## Additional files

**Additional file 1: Table S1.** Compartmental model PK parameters. **Table S2.** PK parameter sourcing. **Table S3.** Scaling EIR by reducing larval habitat. **Table S4.** Correlations of prevalence after MDA campaigns with and without PQ.

**Additional file 2: Figure S1.** Asexual parasite prevalence and distribution of asexual parasitaemia and gametocytaemia under constant annual EIR 36 with a semi-immune population in the absence of intervention. **Figure S2.** Campaign outcome dependence on timing, coverage, and compliance. **Figure S3.** Prevalence 4 months after campaigns with and without primaquine.

#### Competing interests
The authors declare that they have no competing interests.

#### Authors' contributions
JG designed the simulations, performed the analysis, and wrote the manuscript. PE and EAW conceptualized the project. All authors have reviewed and approved the final manuscript.

#### Acknowledgements
The authors thank Bill and Melinda Gates for their active support of this work and their sponsorship through the Global Good Fund.

Received: 14 November 2014 Accepted: 12 March 2015
Published online: 22 March 2015

#### References
1. World Health Organization. World Malaria Report 2013. Geneva: WHO; 2013.
2. Greenwood B. The use of anti-malarial drugs to prevent malaria in the population of malaria-endemic areas. Am J Trop Med Hyg. 2004;70:1–7.
3. Song J, Socheat D, Tan B, Dara P, Deng C, Sokunthea S, et al. Rapid and effective malaria control in Cambodia through mass administration of artemisinin-piperaquine. Malaria J. 2010;9:57.
4. von Seidlein L, Walraven G, Milligan PJ, Alexander N, Manneh F, Deen JL, et al. The effect of mass administration of sulfadoxine-pyrimethamine combined with artesunate on malaria incidence: a double-blind, community-randomized, placebo-controlled trial in The Gambia. Trans R Soc Trop Med Hyg. 2003;97:217–25.
5. White NJ. The role of anti-malarial drugs in eliminating malaria. Malaria J. 2008;7 Suppl 1:S8.
6. Poirot E, Skarbinski J, Sinclair D, Kachur SP, Slutsker L, Hwang J. Mass drug administration for malaria. Cochrane Database Syst Rev. 2013;12, CD008846.
7. El-Sayed B, El-Zaki S-E, Babiker H, Gadalla N, Ageep T, Mansour F, et al. A Randomized Open-Label Trial of Artesunate- Sulfadoxine-Pyrimethamine with or without Primaquine for Elimination of Sub-Microscopic P. falciparum Parasitaemia and Gametocyte Carriage in Eastern Sudan. PLoS ONE. 2007;2:e1311.
8. Shekalaghe SA, Drakeley C, van den Bosch S, Braak ter R, van den Bijllaardt W, Mwanziva C, et al. A cluster-randomized trial of mass drug administration with a gametocytocidal drug combination to interrupt malaria transmission in a low endemic area in Tanzania. Malaria J. 2011;10:247.
9. Yeung S, White NJ. How do patients use antimalarial drugs? A review of the evidence. Trop Med Int Health. 2005;10:121–38.
10. Lawford H, Zurovac D, O'Reilly L, Hoibak S, Cowley A, Munga S, et al. Adherence to prescribed artemisinin-based combination therapy in Garissa and Bunyala districts, Kenya. Malaria J. 2011;10:281.
11. Minzi O, Maige S, Sasi P, Ngasala B. Adherence to artemether-lumefantrine drug combination: a rural community experience six years after change of malaria treatment policy in Tanzania. Malaria J. 2014;13:267.
12. Okell LC, Drakeley CJ, Bousema T, Whitty CJM, Ghani AC. Modelling the impact of artemisinin combination therapy and long-acting treatments on malaria transmission intensity. PLoS Med. 2008;5(11):1617–28.
13. White NJ. Primaquine to prevent transmission of falciparum malaria. Lancet Infect Dis. 2013;13:175–81.
14. Eziefula AC, Bousema T, Yeung S, Kamya M, Owaraganise A, Gabagaya G, et al. Single dose primaquine for clearance of Plasmodium falciparum gametocytes in children with uncomplicated malaria in Uganda: a randomised, controlled, double-blind, dose-ranging trial. Lancet Infect Dis. 2014;14:130–9.
15. Beutler E. The hemolytic effect of primaquine and related compounds: a review. Blood. 1959;14:103–39.
16. Reeve PA, Toaliu H, Kaneko A, Hall JJ, Ganczakowski M. Acute intravascular haemolysis in Vanuatu following a single dose of primaquine in individuals with glucose-6-phosphate dehydrogenase deficiency. J Trop Med Hyg. 1992;95:349–51.
17. Molineaux L, Diebner HH, Eichner M, Collins WE, Jeffery GM, Dietz K. Plasmodium falciparum parasitaemia described by a new mathematical model. Parasitology. 2001;122:379–91.
18. Ross R. The Prevention of Malaria. New York: Dutton; 1910.
19. Eckhoff PA. A malaria transmission-directed model of mosquito life cycle and ecology. Malaria J. 2011;10:303.
20. Eckhoff P. P. falciparum Infection Durations and Infectiousness Are Shaped by Antigenic Variation and Innate and Adaptive Host Immunity in a Mathematical Model. PLoS ONE. 2012;7:e44950.
21. Filipe JAN, Riley EM, Drakeley CJ, Sutherland CJ, Ghani AC. Determination of the Processes Driving the Acquisition of Immunity to Malaria Using a Mathematical Transmission Model. PLoS Comput Biol. 2007;3:e255.
22. Hodel EM, Kay K, Hayes DJ, Terlouw DJ, Hastings IM. Optimizing the programmatic deployment of the anti-malarials artemether-lumefantrine and dihydroartemisinin-piperaquine using pharmacological modelling. Malaria Journal. 2014;13:138.
23. Maude RJ, Socheat D, Nguon C, Saroth P, Dara P, Li G, et al. Optimising strategies for Plasmodium falciparum malaria elimination in Cambodia: primaquine, mass drug administration and artemisinin resistance. PLoS ONE. 2012;7:e37166.
24. Griffin JT, Hollingsworth TD, Okell LC, Churcher TS, White M, Hinsley W, et al. Reducing Plasmodium falciparum Malaria Transmission in Africa: A Model-Based Evaluation of Intervention Strategies. PLoS Med. 2010;7:e1000324.
25. Okell LC, Griffin JT, Kleinschmidt I, Hollingsworth TD, Churcher TS, White MJ, et al. The Potential Contribution of Mass Treatment to the Control of Plasmodium falciparum Malaria. PLoS ONE. 2011;6:e20179.
26. Silal SP, Little F, Barnes KI, White LJ. Towards malaria elimination in Mpumalanga, South Africa: a population-level mathematical modelling approach. Malaria J. 2014;13:297.
27. Johnston GL, Gething PW, Hay SI, Smith DL, Fidock DA. Modeling Within-Host Effects of Drugs on Plasmodium falciparum Transmission and Prospects for Malaria Elimination. PLoS Comput Biol. 2014;10:e1003434.
28. Ezzet F, Mull R, Karbwang J. Population pharmacokinetics and therapeutic response of CGP 56697 (artemether + benflumetol) in malaria patients. Br J Clin Pharmacol. 1998;46:553–61.
29. Ezzet F, van Vugt M, Nosten F, Looareesuwan S, White NJ. Pharmacokinetics and pharmacodynamics of lumefantrine (benflumetol) in acute falciparum malaria. Antimicrob Agents Chemother. 2000;44:697–704.
30. Denis MB, Tsuyuoka R, Lim P, Lindegardh N, Yi P, Top SN, et al. Efficacy of artemether-lumefantrine for the treatment of uncomplicated falciparum malaria in northwest Cambodia. Trop Med Int Health. 2006;11:1800–7.
31. Delves M, Plouffe D, Scheurer C, Meister S, Wittlin S, Winzeler EA, et al. The Activities of Current Antimalarial Drugs on the Life Cycle Stages of Plasmodium: A Comparative Study with Human and Rodent Parasites. PLoS Med. 2012;9:e1001169.
32. Tyner SD, Lon C, Se Y, Bethell D, Socheat D, Noedl H, et al. Ex vivo drug sensitivity profiles of Plasmodium falciparum field isolates from Cambodia and Thailand, 2005 to 2010, determined by a histidine-rich protein-2 assay. Malaria J. 2012;11:198.
33. Tarning J, Ashley EA, Lindegardh N, Stepniewska K, Phaiphun L, Day NPJ, et al. Population Pharmacokinetics of Piperaquine after Two Different Treatment Regimens with Dihydroartemisinin-Piperaquine in Patients with Plasmodium falciparum Malaria in Thailand. Antimicrob Agents Chemother. 2008;52:1052–61.
34. Piyaphanee W, Krudsood S, Tangpukdee N, Thanachartwet W, Silachamroon U, Phophak N, et al. Emergence and clearance of gametocytes in uncomplicated Plasmodium falciparum malaria. Am J Trop Med Hyg. 2006;74:432–5.
35. Denis MB, Davis TME, Hewitt S, Incardona S, Nimol K, Fandeur T, et al. Efficacy and safety of dihydroartemisinin-piperaquine (Artekin) in Cambodian children and adults with uncomplicated falciparum malaria. Clin Infect Dis. 2002;35:1469–76.

36. Mårtensson A, Strömberg J, Sisowath C, Msellem MI, Gil JP, Montgomery SM, et al. Efficacy of artesunate plus amodiaquine versus that of artemether-lumefantrine for the treatment of uncomplicated childhood Plasmodium falciparum malaria in Zanzibar, Tanzania. Clin Infect Dis. 2005;41:1079–86.
37. Piola P, Fogg C, Bajunirwe F, Biraro S, Grandesso F, Ruzagira E, et al. Supervised versus unsupervised intake of six-dose artemether-lumefantrine for treatment of acute, uncomplicated Plasmodium falciparum malaria in Mbarara, Uganda: a randomised trial. Lancet. 2005;365:1467–73.
38. Tshefu AK, Gaye O, Kayentao K, Thompson R, Bhatt KM, Sesay SS, et al. Efficacy and safety of a fixed-dose oral combination of pyronaridine-artesunate compared with artemether-lumefantrine in children and adults with uncomplicated Plasmodium falciparum malaria: a randomised non-inferiority trial. The Lancet. 2010;375:1457–67.
39. Price RN, Uhlemann A-C, van Vugt M, Brockman A, Hutagalung R, Nair S, et al. Molecular and pharmacological determinants of the therapeutic response to artemether-lumefantrine in multidrug-resistant Plasmodium falciparum malaria. Clin Infect Dis. 2006;42:1570–7.
40. Bassat Q, Mulenga M, Tinto H, Piola P, Borrmann S, Menéndez C, et al. Dihydroartemisinin-Piperaquine and Artemether-Lumefantrine for Treating Uncomplicated Malaria in African Children: A Randomised, Non-Inferiority Trial. PLoS ONE. 2009;4:e7871.
41. The Four Artemisinin-Based Combinations (4ABC) Study Group. A Head-to-Head Comparison of Four Artemisinin-Based Combinations for Treating Uncomplicated Malaria in African Children: A Randomized Trial. PLoS Med. 2011;8:e1001119.
42. Smithuis F, Kyaw MK, Phe O, Win T, Aung PP, Oo APP, et al. Effectiveness of five artemisinin combination regimens with or without primaquine in uncomplicated falciparum malaria: an open-label randomised trial. Lancet Infect Dis. 2010;10:673–81.
43. Pukrittayakamee S, Chotivanich K, Chantra A, Clemens R, Looareesuwan S, White NJ. Activities of Artesunate and Primaquine against Asexual- and Sexual-Stage Parasites in Falciparum Malaria. Antimicrob Agents Chemother. 2004;48:1329–34.
44. White NJ, Qiao LG, Qi G, Luzzatto L. Rationale for recommending a lower dose of primaquine as a Plasmodium falciparum gametocytocide in populations where G6PD deficiency is common. Malaria J. 2012;11:418.
45. Eckhoff P. Mathematical models of within-host and transmission dynamics to determine effects of malaria interventions in a variety of transmission settings. Am J Trop Med Hyg. 2013;88:817–27.
46. Tarning J, Kloprogge F, Piola P, Dhorda M, Muwanga S, Turyakira E, et al. Population pharmacokinetics of Artemether and dihydroartemisinin in pregnant women with uncomplicated Plasmodium falciparum malaria in Uganda. Malaria J. 2012;11:293.
47. Na Bangchang K, Karbwang J, Thomas CG, Thanavibul A, Sukontason K, Ward SA, et al. Pharmacokinetics of artemether after oral administration to healthy Thai males and patients with acute, uncomplicated falciparum malaria. Br J Clin Pharmacol. 1994;37:249–53.
48. Mwesigwa J, Parikh S, McGee B, German P, Drysdale T, Kalyango JN, et al. Pharmacokinetics of Artemether-Lumefantrine and Artesunate-Amodiaquine in Children in Kampala, Uganda. Antimicrob Agents Chemother. 2009;54:52–9.
49. Djimdé A, Lefèvre G. Understanding the pharmacokinetics of Coartem. Malaria J. 2009;8 Suppl 1:S4.
50. Tarning J, Rijken MJ, McGready R, Phyo AP, Hanpithakpong W, Day NPJ, et al. Population Pharmacokinetics of Dihydroartemisinin and Piperaquine in Pregnant and Nonpregnant Women with Uncomplicated Malaria. Antimicrob Agents Chemother. 2012;56(4):1997–2007.
51. Nguyen DVH, Nguyen QP, Nguyen ND, Le TTT, Nguyen TD, Dinh DN, et al. Pharmacokinetics and Ex Vivo Pharmacodynamic Antimalarial Activity of Dihydroartemisinin-Piperaquine in Patients with Uncomplicated Falciparum Malaria in Vietnam. Antimicrob Agents Chemother. 2009;53:3534–7.
52. Hung T-Y, Davis TME, Ilett KF, Karunajeewa H, Hewitt S, Denis MB, et al. Population pharmacokinetics of piperaquine in adults and children with uncomplicated falciparum or vivax malaria. Br J Clin Pharmacol. 2004;57:253–62.
53. Moore BR, Salman S, Benjamin J, Page-Sharp M, Robinson LJ, Waita E, et al. Pharmacokinetic Properties of Single-Dose Primaquine in Papua New Guinean Children: Feasibility of Abbreviated High-Dose Regimens for Radical Cure of Vivax Malaria. Antimicrob Agents Chemother. 2013;58:432–9.
54. Na Bangchang K, Songsaeng W, Thanavibul A, Choroenlarp P, Karbwang J. Pharmacokinetics of primaquine in G6PD deficient and G6PD normal patients with vivax malaria. Trans R Soc Trop Med Hyg. 1994;88:220–2.
55. Binh VQ, Chinh NT, Thanh NX, Cuong BT, Quang NN, Dai B, et al. Sex Affects the Steady-State Pharmacokinetics of Primaquine but Not Doxycycline in Healthy Subjects. Am J Trop Med Hyg. 2009;81:747–53.
56. Kim Y-R, Kuh H-J, Kim M-Y, Kim Y-S, Chung W-C, Kim S-I, et al. Pharmacokinetics of primaquine and carboxyprimaquine in Korean patients with vivax malaria. Arch Pharm Res. 2004;27:576–80.
57. Bhatia SC, Saraph YS, Revankar SN, Doshi KJ, Bharucha ED, Desai ND, et al. Pharmacokinetics of primaquine in patients with P. vivax malaria. Eur J Clin Pharmacol. 1986;31:205–10.
58. Edwards G, McGrath CS, Ward SA, Supanaranond W, Pukrittayakamee S, Davis TM, et al. Interactions among primaquine, malaria infection and other antimalarials in Thai subjects. Br J Clin Pharmacol. 1993;35:193–8.
59. Hodel EMS, Guidi M, Zanolari B, Mercier T, Duong S, Kabanywanyi AM, et al. Population pharmacokinetics of mefloquine, piperaquine and artemether-lumefantrine in Cambodian and Tanzanian malaria patients. Malaria J. 2013;12:235.
60. Wagner JG. Fundamentals of Clinical Pharmacokinetics. Hamilton: Drug Intelligence Publications; 1975.
61. World Health Organization. Guidelines for the Treatment of Malaria. World Health Organization; 2010.
62. Simpson JA, Watkins ER, Price RN, Aarons L, Kyle DE, White NJ. Mefloquine pharmacokinetic-pharmacodynamic models: implications for dosing and resistance. Antimicrob Agents Chemother. 2000;44:3414–24.
63. Prinz H. Hill coefficients, dose–response curves and allosteric mechanisms. J Chem Biol. 2009;3:37–44.
64. Adjalley SH, Johnston GL, Li T, Eastman RT, Ekland EH, Eappen AG, et al. Quantitative assessment of Plasmodium falciparum sexual development reveals potent transmission-blocking activity by methylene blue. Proc Natl Acad Sci USA. 2011;108:E1214–23.
65. Tarning J, Zongo I, Somé FA, Rouamba N, Parikh S, Rosenthal PJ, et al. Population Pharmacokinetics and Pharmacodynamics of Piperaquine in children with Uncomplicated Falciparum Malaria. Clin Pharmacol Therapeut. 2009;91:497–505.
66. The WorldWide Antimalarial Resistance Network (WWARN) DP Study Group. The Effect of Dosing Regimens on the Antimalarial Efficacy of Dihydroartemisinin-Piperaquine: A Pooled Analysis of Individual Patient Data. PLoS Med. 2013;10:e1001564.
67. Hodel EM, Kay K, Hayes DJ, Terlouw DJ, Hastings IM. Optimizing the programmatic deployment of the anti-malarials artemether-lumefantrine and dihydroartemisinin-piperaquine using pharmacological modelling. Malaria J. 2014;13:138.
68. Falade C, Makanga M, Premji Z, Ortmann C-E, Stockmeyer M, de Palacios PI. Efficacy and safety of artemether–lumefantrine (Coartem®) tablets (six-dose regimen) in African infants and children with acute, uncomplicated falciparum malaria. Trans R Soc Trop Med Hyg. 2005;99:459–67.
69. Gu W, Killeen GF, Mbogo CM, Regens JL, Githure JI, Beier JC. An individual-based model of Plasmodium falciparum malaria transmission on the coast of Kenya. Trans R Soc Trop Med Hyg. 2003;97:43–50.
70. Babiker HA, Schneider P, Reece SE. Gametocytes: insights gained during a decade of molecular monitoring. Trends Parasitol. 2008;24:525–30.
71. Halliday KE, Okello G, Turner EL, Njagi K, Mcharo C, Kengo J, et al. Impact of Intermittent Screening and Treatment for Malaria among School Children in Kenya: A Cluster Randomised Trial. PLoS Med. 2014;11:e1001594.
72. von Seidlein L. The failure of screening and treating as a malaria elimination strategy. PLoS Med. 2014;11:e1001595.
73. Tiono AB, Ouédraogo A, Ogutu B, Diarra A, Coulibaly S, Gansané A, et al. A controlled, parallel, cluster-randomized trial of community-wide screening and treatment of asymptomatic carriers of Plasmodium falciparum in Burkina Faso. Malaria J. 2013;12:79.